\documentstyle[aps,amssymb]{revtex}

\begin{document}
\title{Self-dual teleparallel gravity and the positive energy theorem}
\author{G. Y.\ Chee}
\address{Physics Department, Liaoning Normal University, \\
Dalian, 116029, China\\
Institute of Applied Mathematics, Academy of Mathematics \\
and System Science, Chinese Academy of Science, \\
PO Box 2734, Beijing, 100080, China}
\maketitle

\begin{abstract}
A self-dual and anti-self-dual decomposition of the teleparallel gravity is
carried out and the self-dual Lagrangian of the teleparallel gravity which
is equivalent to the Ashtekar Lagrangian in vacuum is obtained. Its
Hamiltonian formulation and the constraint analysis are developed. Starting
from Witten's equation Nester's gauge condition is derived directly and a
new expression of the boundary term is obtained. \ Using this expression and
Witten's identity the proof of the positive energy theorem by Nester et al
is extended to a case including momentum.

PACS number(s): 04.50.+h, 04.20.Gz, 04.20.Fy
\end{abstract}

\section{Introduction}

Recently, as a description of gravity equivalent to general relativity the
teleparallel gravity has attracted renewed attention [1-5] owing to many
salient features of it. First of all, the teleparallel gravity can be
regarded as a translational gauge theory [1, 2, 4, 6], which make it
possible to unify gravity with other kinds of interactions in the gauge
theory framework in which the elementary interactions are described by a
connection defined on some principal fiber bundle. In this direction
interesting developments [7] have been achieved in the context of Ashtekar
variables [8]. However, when dealing with supergravity and in the search for
the construction of a unified model for the fundamental interactions,
another usual route [9] is to consider Kaluza-klein type models [10] of
supergravity as candidates. In this approach the fields describing the
fundamental interactions (including gravity) correspond to different pieces
of the pseudo Riemannian \ metric characterizing a higher dimensional
spacetime. This approach are considered as quite different from the former
and the relation between them has note been very clear. It is noteworthy
that some teleparallel equivalents of the Kaluza-Klein theory and
non-Abelian Kaluza-Klein theory are developed [11], which gives us new
perspectives for the study of unified theories.

Another advantage of the teleparallel gravity concerns energy-momentum, its
representation, positivity and localization [1, 2, 5]. Because of its
simplicity and transparency the teleparallel gravity seems to be much more
appropriate than general relativity to deal with the problem of the
gravitational energy-momentum. It is proved that [12, 1, 2, 5] in the
teleparallel gravity there exists a gravitational energy-momentum tensor
which is covariant under general coordinate transformations and global
Lorentz transformations.

Attempts at identifying an energy-momentum density for gravity in the
context of general relativity lead only to various energy-momentum complexes
which are pseudotensors and then a new quasilocal approach which can be
traced back to the early work of Penrose [13] has been proposed and become
widely accepted [5, 14]. According to this approach a quasilocal
energy-momentum can be obtained from the Hamiltonian. Every energy-momentum
pseudotensor is associated with a legitimated Hamiltonian boundary term. In
terms of teleparallel gravity a geometrically natural proof of the
positivity of the gravitational energy is obtained [5] by choosing a maximal
surface and a vanishing shift.

Nester and his coworkers have found a four-spinor formulation of the
teleparallel gravity [5].This formulation has several virtues, in particular
it gives a four-covariant Hamiltonian, shows that total four-momentum is
future timelike and can be evaluated on a spacelike surface extending to
future null infinity thereby showing that the Bondi four-momentum also is
future timelike. It is suggested to generalizing this formulation to
self-dual representations. The chiral \ Lagrangian formulation of general
relativity employing two-component spinors has been introduced [15]. Usually
the teleparallel gravity is viewed as a equivalent of general relativity. It
is well known, as a self-dual formulation of general relativity Ashtekar's
theory opens new avenues to quantum gravity and plays a important role in
the development of modern gravitational theory. It is shown \ that a lot of
gauge theories, gravity and supergravity theories have their self-dual
partners [16-18], a question naturally arises whether there is a self-dual
teleparallel gravitational theory which equivalent to Ashtekar's theory. If
it exists, can it gives us some new perspectives? In this paper a two-spinor
formulation of the teleparallel gravity which is a self-dual representation
of the teleparallel gravity and equivalent to the Ashtekar theory [8] will
be developed.

In the proof of the positive energy theorem [19] Witten propose a spatial
Dirac equation. In the last two decades people have been trying to
understand the meaning of this equation and its solutions. In terms of a
orthonormal frame Nester gave another proof of the positive energy theorem
[5] using teleparallel geometry under a special gauge in which the shift
vanishes. Some authors found the relation between the orthonormal frame
(triad) and the Witten equation [20]. In this paper Nester's gauge condition
will be derived from Witten's equation. Furthermore a proof of the positive
energy theorem different from Nester's by a nonvanishing shift and without
maximal surface will be suggested.

In section 2 the self-dual and anti-self-dual decomposition of the
Lagrangian of the teleparallel gravity is carried out and the self-dual
teleparallel Lagrangian is given. In section 3 the Hamiltonian formulation
and the constraint algebra of the teleparallel gravity is built up. Its
boundary term is just the self-dual part of Nester's boundary term. In
section 4, using Witten's equation Nester's gauge condition is derived and a
new expression of the boundary term is obtained. Using this expression and
the Witten identity a proof of the positive energy theorem is shown under a
deferent gauge condition from the one of Nester in section 5. Finally,
section 6 is devoted to some conclusions.

\section{Self-dual and anti-self-dual decomposition of the Lagrangian of the
teleparallel gravity}

We start with a common relation between the tetrad $e^I\!_\mu $, the spin
connection $\omega _\mu \!^I\!_J$, and the affine connection $\stackrel{%
\circ }{\Gamma }^\rho \!_{\nu \mu }\;$[1, 21] 
\begin{equation}
\partial _\mu e^{I{}}\!_\nu +\omega _\mu \!^I\!_Je^J\!_\nu -\stackrel{\circ 
}{\Gamma }^\rho \!_{\mu \nu }e^I\!_\rho =0,
\end{equation}
where $I,J,\ldots =0,1,2,3$ are the internal indices and $\mu ,\nu ,\ldots
=0,1,2,3\;$are the spacetime indices. Defining the Weitzenbok connection
[1,21] 
\begin{equation}
\Gamma ^\rho {}_{\mu \nu }=e\!_I\!^\rho \partial _\mu e^{I{}}\!_\nu ,
\end{equation}
then (1) leads to 
\begin{eqnarray}
\Gamma ^\rho {}_{\mu \nu } &=&\stackrel{\circ }{\Gamma }^\rho \!_{\mu \nu
}-\omega _\mu \!^I\!_Je_I\!^\rho e^J\!_\nu  \nonumber \\
&=&\stackrel{\circ }{\Gamma }^\rho \!_{\mu \nu }-\omega _\mu \!^\rho \!_\nu 
\nonumber \\
&=&\{_\mu \!^\rho \!_\nu \}+\stackrel{\circ }{K}^\rho \!_{\mu \nu }-\omega
_\mu \!^\rho \!_\nu ,
\end{eqnarray}
where 
\begin{equation}
\omega _\mu \!^\rho \!_\nu =\omega _\mu \!^I\!_Je_I\!^\rho e^J\!_\nu ,
\end{equation}
and $\{_\mu \!^\rho \!_\nu \}$,$\;\stackrel{\circ }{K}^\rho \!_{\mu \nu }$
is the Christoffel connection and the affine contortion, respectively. By
introducing the Weitzenbok torsion 
\begin{equation}
T^\rho {}_{\mu \nu }=\Gamma ^\rho {}_{\mu \nu }-\Gamma ^\rho {}_{\nu \mu },
\end{equation}
and the Weitzenbok$\ $contortion [1,22]

\begin{equation}
K^\rho {}_{\mu \nu }\!=\frac 12(T^\rho \!_{\nu \mu }+T_\mu \!^\rho \!_\nu
+T_\nu \!^\rho \!_\mu ),
\end{equation}
we can obtain from (3) 
\begin{equation}
T^\rho {}_{\mu \nu }=\stackrel{\circ }{T}^\rho \!_{\mu \nu }-\omega _\mu
\!^\rho \!_\nu +\omega _\nu \!^\rho \!_\mu ,
\end{equation}
and 
\begin{equation}
K^\rho \!_{\mu \nu }=\stackrel{\circ }{K}^\rho \!_{\mu \nu }+\omega _{\!\mu
\nu }\!^\rho ,
\end{equation}
where 
\begin{equation}
\stackrel{\circ }{T}^\rho \!_{\mu \nu }=2\stackrel{\circ }{\Gamma }^\rho
\!_{[\mu \nu ]}=\stackrel{\circ }{\Gamma }^\rho \!_{\mu \nu }-\stackrel{%
\circ }{\Gamma }^\rho \!_{\nu \mu },
\end{equation}
is the affine torsion and 
\begin{equation}
\stackrel{\circ }{K}^\rho \!_{\mu \nu }=\frac 12(\stackrel{\circ }{T}^\rho
\!_{\mu \nu }+\stackrel{\circ }{T}_\mu \!^\rho \!_\nu +\stackrel{\circ }{T}%
_\nu \!^\rho \!_\mu ),
\end{equation}
is the affine contortion [22].

In this paper we concentrate on the case of vanishing affine torsion 
\begin{equation}
\stackrel{\circ }{T}^\rho \!_{\mu \nu }=0,
\end{equation}
and then 
\[
\stackrel{\circ }{\Gamma }^\rho \!_{\mu \nu }=\{_\mu \!^\rho \!_\nu \} 
\]
as in the usually general relativity, we have 
\begin{equation}
\stackrel{\circ }{K}^\rho \!_{\mu \nu }=0,
\end{equation}
and then 
\begin{equation}
K^\rho \!_{\mu \nu }=\omega _{\mu \!\nu }\!^\rho .
\end{equation}

As a result (3) reads 
\begin{eqnarray}
\Gamma ^\rho {}_{\mu \nu } &=&\{_\mu \!^\rho \!_\nu \}+K^\rho \!_{\mu \nu } 
\nonumber \\
&=&\stackrel{\circ }{\Gamma }^\rho \!_{\mu \nu }+K^\rho \!_{\mu \nu },
\end{eqnarray}
and we are led to the theories given by Hayashi, Shirafuji [22], de Andrade,
Guillen and Pereira [1] which are equivalent to general relativity and
called theories of teleparallel gravity. In these theories the curvature of
the Weitzenbok connection vanishes: 
\begin{equation}
R^\rho \!_{\sigma \mu \nu }=\partial _\mu \Gamma ^\rho {}_{\sigma \nu
}-\partial _\nu \Gamma ^\rho {}_{\sigma \mu }+\Gamma ^\rho {}_{\tau \mu
}\Gamma ^\tau {}_{\sigma \nu }-\Gamma ^\rho {}_{\tau \nu }\Gamma ^\tau
{}_{\sigma \mu }\equiv 0,
\end{equation}
while the curvature of the Christoffel connection 
\begin{equation}
\stackrel{\circ }{R}^\rho \!_{\sigma \mu \nu }=\partial _\mu \{_\sigma
\!^\rho \!_\nu \}-\partial _\nu \{_\sigma \!^\rho \!_\mu \}+\{_\tau \!^\rho
\!_\mu \}\{_\sigma \!^\tau \!_\nu \}-\{_\tau \!^\rho \!_\nu \}\{_\sigma
\!^\tau \!_\mu \}
\end{equation}
does not. For these theories one can say that the spacetime is a Weitzenbok
spacetime with respect to the Cartan connection or a Riemann spacetime with
respect to the Christoffel connection.

According to [1], the Lagrangian of the gravitational field can be chosen as

\begin{equation}
{\cal L}_{G}=-\frac{^{(4)}e}{2}S^{\rho \mu \nu }T_{\rho \mu \nu },
\end{equation}
where $^{(4)}e=\det (e^{I}\!_{\mu })$, and 
\begin{equation}
S^{\rho \mu \nu }=\frac{1}{2}(K^{\mu \nu \rho }-g^{\rho \nu }T^{\sigma \mu
}\!_{\sigma }+g^{\rho \mu }T^{\sigma \nu }\!_{\sigma }).
\end{equation}

From (7) and (11) we can obtain 
\begin{equation}
T^\rho {}{}_{\mu \nu }=\omega _\nu \!^\rho \!_\mu -\omega _\mu \!^\rho
\!_\nu ,
\end{equation}
and 
\begin{equation}
S_\mu \!^{\nu \rho }=\frac 12(\omega ^\rho \!_\mu \!^\nu -\delta _\mu ^\rho
\omega _\sigma \!^{\sigma \nu }+\delta _\mu ^\nu \omega _\sigma \!^{\sigma
\rho }),
\end{equation}
which lead to 
\begin{equation}
S^{\rho \mu \nu }T_{\rho \mu \nu }=\omega _\rho \!^{\mu \nu }\omega _\mu
\!^\rho \!_\nu -\omega _\rho \!^{\rho \nu }\omega _\mu \!^\mu \!_\nu ,
\end{equation}
and 
\begin{equation}
S_\mu \!^{\nu \rho }T^\mu \!_{\nu \lambda }=\frac 12\omega ^\rho \!_\mu
\!^\nu (\omega _\lambda \!^\mu \!_\nu -\omega _\nu \!^\mu \!_\lambda )-\frac %
12\omega _\mu \!^{\mu \nu }(\omega _\lambda \!^\rho \!_\nu -\omega _\nu
\!^\rho \!_\lambda )-\frac 12\omega _\mu \!^{\mu \rho }\omega _\nu \!^\nu
\!_\lambda .
\end{equation}

Using (13), (6) and (5) we can compute 
\begin{eqnarray}
\omega _{\mu \nu \rho } &=&K_{\rho \mu \nu }=\frac{1}{2}(T_{\mu \rho \nu
}+T_{\nu \rho \mu }+T_{\rho \mu \nu })  \nonumber \\
&=&(e_{I\rho }\partial _{\lbrack \mu }e^{I}\!_{\nu ]}+e_{I\mu }\partial
_{\lbrack \rho }e^{I}\!_{\nu ]}+e_{I\nu }\partial _{\lbrack \rho
}e^{I}\!_{\mu ]}),
\end{eqnarray}
\begin{equation}
\omega ^{\nu \mu \rho }=g^{\nu \tau }g^{\rho \sigma }e_{I}\!^{\mu }\partial
_{\lbrack \sigma }e^{I}\!_{\tau ]}+g^{\mu \lambda }g^{\rho \sigma
}e_{I}\!^{\nu }\partial _{\lbrack \sigma }e^{I}\!_{\lambda ]}+g^{\mu \lambda
}g^{\nu \tau }e_{I}\!^{\rho }\partial _{\lbrack \lambda }e^{I}\!_{\tau ]},
\end{equation}
and 
\begin{eqnarray}
S^{\rho \mu \nu }T_{\rho \mu \nu } &=&\omega _{\mu \nu \rho }\omega ^{\nu
\mu \rho }-\omega _{\mu }\!^{\mu \rho }\omega _{\nu }\!^{\nu }\!_{\rho } 
\nonumber \\
&=&\frac{1}{4}T_{\rho \mu \nu }T^{\rho \mu \nu }+\frac{1}{2}T_{\rho \mu \nu
}T^{\nu \mu \rho }+T_{\rho \mu }\!\!^{\rho }T^{\nu \!}\!_{\nu }\!^{\mu }.
\end{eqnarray}
Then the Lagrangian (17) can be written as 
\begin{eqnarray}
{\cal L}_{G} &=&-\frac{^{(4)}e}{8}(\eta _{IJ}g^{\mu \lambda }g^{\nu \tau
}+2^{(4)}e_{I}\!^{\tau (4)}e_{J}\!^{\nu }g^{\mu \lambda
}-4^{(4)}e_{I}\!^{\mu (4)}e_{J}\!^{\lambda }g^{\nu \tau })T^{I}\!_{\mu \nu
}T^{J}\!_{\lambda \tau }  \nonumber \\
&=&-\frac{^{(4)}e}{8}(T_{\mu \nu \lambda }T^{\mu \nu \lambda }+2T_{\mu \nu
\lambda }T^{\lambda \nu \mu }-4T^{\mu }{}_{\mu \lambda }T^{\nu }{}_{\nu
}{}^{\lambda }),
\end{eqnarray}
which is equivalent to the Einstein Lagrangian.

In terms of two-spinors the connection $\omega _{\mu \nu }{}^\rho =\omega
_{AA^{\prime }BB^{\prime }}{}^{CC^{\prime }}$ can be decomposed into two
parts: 
\begin{eqnarray}
\omega _{AA^{\prime }BB^{\prime }}{}^{CC^{\prime }} &=&\omega _{AA^{\prime
}B}{}^C\epsilon ^{C^{\prime }}{}_{B^{\prime }}+\overline{\omega }%
_{AA^{\prime }B^{\prime }}{}^{C\prime }\epsilon ^C{}_B  \nonumber \\
&=&\omega _{AA^{\prime }BB^{\prime }}^{+}{}^{CC^{\prime }}+\omega
_{AA^{\prime }BB^{\prime }}^{-}{}^{CC^{\prime }},
\end{eqnarray}
where 
\begin{equation}
\omega _{AA^{\prime }BB^{\prime }}^{+}{}^{CC^{\prime }}=\omega _{AA^{\prime
}B}{}^C\epsilon ^{C^{\prime }}{}_{B^{\prime }},
\end{equation}
and 
\begin{equation}
\omega _{AA^{\prime }BB^{\prime }}^{-}{}^{CC^{\prime }}=\overline{\omega }%
_{AA^{\prime }B^{\prime }}{}^{C\prime }\epsilon ^C{}_B,
\end{equation}
is the self-dual and the anti-self-dual part of the connection $\omega
_{AA^{\prime }BB^{\prime }}{}^{CC^{\prime }}$

($A,B,\ldots =0,1;A^{\prime },B^{\prime },\ldots =0^{\prime },1^{\prime }$),
respectively. Using these results we obtain 
\begin{eqnarray*}
T_{(c)\mu \nu \lambda }T_{(c)}^{\mu \nu \lambda } &=&T_{AA^{\prime
}BB^{\prime }CC^{\prime }}T^{AA^{\prime }BB^{\prime }CC^{\prime }} \\
&=&4\omega _{AA^{\prime }BC}\omega ^{AA^{\prime }BC}+4\overline{\omega }%
_{AA^{\prime }B^{\prime }C^{\prime }}\overline{\omega }^{AA^{\prime
}B^{\prime }C^{\prime }} \\
&&-2\omega _{AB^{\prime }CB}\omega ^{BB^{\prime }AC}-2\overline{\omega }%
_{BA^{\prime }C^{\prime }B^{\prime }}\overline{\omega }^{BB^{\prime
}A^{\prime }C^{\prime }} \\
&&+4\omega _{AA^{\prime }}{}^{AB}\overline{\omega }_{BB^{\prime
}}{}^{B^{\prime }A^{\prime }},
\end{eqnarray*}
\begin{eqnarray*}
T_{(c)\mu \nu \lambda }T_{(c)}^{\lambda \nu \mu } &=&-2\omega _{AA^{\prime
}BC}\omega ^{AA^{\prime }BC}-2\overline{\omega }_{AA^{\prime }B^{\prime
}C^{\prime }}\overline{\omega }^{AA^{\prime }B^{\prime }C^{\prime }} \\
&&+3\omega _{AB^{\prime }CB}\omega ^{BB^{\prime }AC}+3\overline{\omega }%
_{BA^{\prime }C^{\prime }B^{\prime }}\overline{\omega }^{BB^{\prime
}A^{\prime }C^{\prime }} \\
&&-6\omega _{AA^{\prime }}{}^{AB}\overline{\omega }_{BB^{\prime
}}{}^{B^{\prime }A^{\prime }},
\end{eqnarray*}
and 
\[
T_{(c)}^\mu {}_{\mu \lambda }T_{(c)}^\nu {}_\nu {}^\lambda =\omega
_{AA^{\prime }}{}^{AC}\omega ^{AB^{\prime }}{}_{AC}+\overline{\omega }%
_{AA^{\prime }}{}^{A^{\prime }C^{\prime }}\overline{\omega }^{AB^{\prime
}}{}_{B^{\prime }C^{\prime }}-2\omega _{AA^{\prime }}{}^{AB}\overline{\omega 
}_{BB^{\prime }}{}^{B^{\prime }A^{\prime }}. 
\]
The Lagrangian ${\cal L}_G$ takes the form 
\begin{eqnarray*}
{\cal L}_G &=&-\frac{^{(4)}\sigma }4(4\omega _{AB^{\prime }CB}\omega
^{BB^{\prime }AC}+4\overline{\omega }_{BA^{\prime }C^{\prime }B^{\prime }}%
\overline{\omega }^{BB^{\prime }A^{\prime }C^{\prime }} \\
&&-4\omega _{AA^{\prime }}{}^{AC}\omega ^{AB^{\prime }}{}_{AC}-4\overline{%
\omega }_{AA^{\prime }}{}^{A^{\prime }C^{\prime }}\overline{\omega }%
^{AB^{\prime }}{}_{B^{\prime }C^{\prime }}) \\
&=&^{(4)}\sigma (\omega _{AA^{\prime }}{}^{AC}\omega ^{BA^{\prime }}{}_{BC}+%
\overline{\omega }_{AA^{\prime }}{}^{A^{\prime }C^{\prime }}\overline{\omega 
}^{AB^{\prime }}{}_{B^{\prime }C^{\prime }} \\
&&-\omega _{AB^{\prime }CB}\omega ^{BB^{\prime }AC}-\overline{\omega }%
_{BA^{\prime }C^{\prime }B^{\prime }}\overline{\omega }^{BB^{\prime
}A^{\prime }C^{\prime }}),
\end{eqnarray*}
and splits into two parts: 
\[
{\cal L}_G={\cal L}_G^{+}+{\cal L}_G^{-}, 
\]
where 
\[
{\cal L}_G^{+}=^{(4)}\sigma (\omega _{AA^{\prime }}{}^{AC}\omega
^{BA^{\prime }}{}_{BC}-\omega _{AB^{\prime }CB}\omega ^{BB^{\prime }AC}), 
\]
and 
\[
{\cal L}_G^{-}=^{(4)}\sigma (\overline{\omega }_{AA^{\prime }}{}^{A^{\prime
}C^{\prime }}\overline{\omega }^{AB^{\prime }}{}_{B^{\prime }C}-\overline{%
\omega }_{BA^{\prime }C^{\prime }B^{\prime }}\overline{\omega }^{BB^{\prime
}A^{\prime }C^{\prime }}), 
\]
is the self-dual part and the anti-self-dual part of ${\cal L}_G$ with the
determinant $^{(4)}\sigma $ of the inverse SL(2,C) soldering form $\sigma
_\mu {}^{AA^{\prime }}$ on the spacetime manifold $M$.

Since ${\cal L}_{G}$ consists of two invariant parts depending on $\omega
_{AA^{\prime }B}{}^{C}$ and $\overline{\omega }_{AA^{\prime }B^{\prime
}}{}^{C\prime }$ respectively, we can choose the self-dual part ${\cal L}%
_{G}^{+}$ as the Lagrangian which is the equivalent of the Ashtekar
Lagrangian [8].

\section{The Hamiltonian formulation of the self-dual teleparallel gravity}

In order to obtain the Hamiltonian of the theory a foliation in the
spacetime manifold $M$ should be introduced. Assuming that $M=\Sigma \times
R $ for some space-like Manifold $\Sigma $, we can choose a time function $t$
with nowhere vanishing gradient $(dt)_\mu $ such that each $t=const$ surface 
$\Sigma _t$ is diffeomorphic to $\Sigma $. Introduce a time flow vector $%
t^\mu $ satisfying $t^\mu (dt)_\mu =1$, we can decompose it perpendicular
and parallel to $\Sigma _t$: $t^\mu =Nn^\mu +N^\mu $, where $n^\mu $ is the
time-like normal at each point of $\Sigma _t$ and $N$, $N^\mu $ are the
lapse function and the shift vector , respectively. The spacetime metric $%
g_{\mu \nu }$ introduces a spatial metric $q_{\mu \nu }$ on each $\Sigma
_t\; $by the formula 
\begin{equation}
q_{\mu \nu }=g_{\mu \nu }+n_\mu n_\nu .
\end{equation}
In the two-spinor formalism the unit normal vector $n^\mu =n^{AA^{\prime }}$
defines an isomorphism from the space of primed spinors to the space of
unprimed spinors [24, 8]: 
\[
\xi ^A=\sqrt{2}n^{AA^{\prime }}\xi _{A^{\prime }}, 
\]
\begin{equation}
\omega _{ABCD}=\sqrt{2}n_A{}^{A^{\prime }}\omega _{AA^{\prime }CD},
\end{equation}
\begin{equation}
n^{AB}=\sqrt{2}n^{BA^{\prime }}n^A{}_{A^{\prime }}=\frac 1{\sqrt{2}}\epsilon
^{AB}.
\end{equation}
In this formalism (30) reads 
\begin{equation}
g^{ABCD}=q^{ABCD}+n^{AB}n^{CD},
\end{equation}
or 
\begin{equation}
\epsilon ^{AC}\epsilon ^{BD}=-\epsilon ^{A(C}\epsilon ^{D)B}+\frac 12%
\epsilon ^{AB}\epsilon ^{CD}.
\end{equation}
Using these results and decomposing $\omega _{ABCD}$ into its symmetry part $%
\omega _{(AB)CD}$ and skew-symmetry part $\omega _{[AB]CD}$: 
\begin{equation}
\omega _{ABCD}=\omega _{(AB)CD}+\omega _{[AB]CD},
\end{equation}
the Lagrangian ${\cal L}_G^{+}$ can be written as 
\[
{\cal L}_G^{+}=^{(4)}\sigma [\omega _{(AB)}{}^{AC}\omega ^{DB}{}_{DC}-\omega
_{(AB)CD}\omega ^{CBAD}-\sqrt{2}\omega _{\perp CD}\omega ^{(CE)}{}_E{}^D], 
\]
where 
\begin{equation}
\omega _{\perp CD}=n^{AB}\omega _{ABCD}.
\end{equation}
From (1) one gets 
\begin{equation}
\omega _\mu \!^I\!_\nu =\omega _\mu \!^I\!_Je^J\!_\nu =-\partial _\mu
e^{I{}}\!_\nu +\Gamma ^\rho \!_{\mu \nu }e^I\!_\rho :=-\stackrel{\circ }{%
\nabla }_\mu e^{I{}}\!_\nu ,
\end{equation}
where $\stackrel{\circ }{\nabla }_\mu $ is the affine covariant derivative
which is just the Christoffel covariant derivative [1] since we have assumed
the vanishing affine torsion. In the two-spinor formalism (37) reads 
\[
\omega _{CDAB}=-\zeta _{aA}\stackrel{\circ }{\nabla }_{CD}\zeta ^a{}_B. 
\]
Using the relation 
\[
n^{AB}=\frac 1N(t^{AB}-N^{AB}), 
\]
one gets 
\begin{equation}
\omega _{\perp CD}=-\frac 1N\zeta _C{}^b\stackrel{\cdot }{\zeta _{bD}}-\frac %
1NN^{AB}\omega _{ABCD},
\end{equation}
where 
\begin{equation}
\stackrel{\cdot }{\zeta _{bD}}=t^{AB}\stackrel{\circ }{\nabla }_{AB}\zeta
_{bD}.
\end{equation}
The Lagrangian ${\cal L}_G^{+}$ becomes 
\begin{eqnarray*}
{\cal L}_G^{+} &=&^{(4)}\sigma [\omega _{(AB)}{}^{AC}\omega
^{(DB)}{}_{DC}-\omega _{(AB)CD}\omega ^{(CB)AD}] \\
&&-^{(4)}\sigma [\omega _{(AB)}{}^{AC}\zeta _D{}^a\stackrel{\circ }{\nabla }%
^{[DB]}\zeta _{aC}-\omega _{(AB)CD}\zeta ^{Aa}\stackrel{\circ }{\nabla }%
^{[CB]}\zeta _a{}^D] \\
&&+^{(4)}\sigma \frac{\sqrt{2}}N(\zeta _C{}^b\stackrel{\cdot }{\zeta _{bD}}%
+N^{AB}\omega _{ABCD})\omega ^{(CE)}{}_E{}^D.
\end{eqnarray*}
The second term can be rewritten 
\begin{eqnarray*}
&&\omega _{(AB)}{}^{AC}\zeta _D{}^a\stackrel{\circ }{\nabla }^{[DB]}\zeta
_{aC}-\omega _{(AB)CD}\zeta ^{Aa}\stackrel{\circ }{\nabla }^{[CB]}\zeta
_a{}^D \\
&=&-\frac{\sqrt{2}}N\omega _{(AB)}{}^{AC}(\zeta ^{Ba}\stackrel{\cdot }{\zeta 
}_{aC}+N^{EF}\omega _{EF}{}{}^B{}_C)
\end{eqnarray*}
and then one obtain

\begin{eqnarray}
{\cal L}_G^{+} &=&N\sigma [\omega _{(AB)}{}^{AC}\omega ^{(DB)}{}_{DC}-\omega
_{(AB)CD}\omega ^{(CB)AD}]  \nonumber \\
&&+2\sqrt{2}\sigma (\zeta _C{}^b\stackrel{\cdot }{\zeta _{bD}}+N^{AB}\omega
_{ABCD})\omega ^{(CE)}{}_E{}^D,
\end{eqnarray}
where 
\begin{equation}
\sigma =\frac{^{(4)}\sigma }N=\det \sigma _\mu {}^{AB}=\frac 1N\sqrt{-g}.
\end{equation}
The canonical momentum conjugate to $\stackrel{.}{\zeta _{bD}}$ is 
\begin{eqnarray}
\widetilde{p}^{bD} &=&\frac{\partial {\cal L}_G^{+}}{\partial \stackrel{%
\cdot }{\zeta _{bD}}}  \nonumber \\
&=&2\sigma \sqrt{2}\zeta _C{}^b\omega ^{(CE)}{}_E{}^D.
\end{eqnarray}
Here $\omega _{(AB)}{}^{CD}$ is just Ashtekar's variable{\bf \ }which
appears in the canonical momentum conjugate $\widetilde{p}^{bD}$ to $\zeta
_{bD}$ and is related to $\widetilde{p}^{bD}$ by 
\begin{equation}
\omega ^{(AC)}{}_C{}^B=-\frac 1{2\sigma \sqrt{2}}\zeta _a{}^A\widetilde{p}%
^{aB}.
\end{equation}

The gravitational Hamiltonian can be computed 
\begin{eqnarray*}
{\cal H}_G &=&\widetilde{p}^{bD}\stackrel{\cdot }{\zeta _{bD}}-{\cal L}_G^{+}
\\
&=&\sigma [N{\cal H}_{\perp }+N^{AB}{\cal H}_{AB}+\stackrel{\circ }{\nabla }%
_{(AB)}B^{AB}],
\end{eqnarray*}
where 
\begin{eqnarray}
{\cal H}_{\perp } &=&\omega _{(AB)CD}\omega ^{(AB)CD}-\frac 1{\sqrt{2}}%
\omega _{(AB)}{}^{AC}\zeta _b{}^B\widetilde{p}^b{}_C  \nonumber \\
&&+\frac 1{2\sqrt{2}}\stackrel{\circ }{\nabla }_{(AB)}(\zeta _b{}^B%
\widetilde{p}^{bA})+\frac 1{2\sqrt{2}}\zeta _b{}^B\widetilde{p}^{bA}%
\stackrel{\circ }{\nabla }_{(AB)}\ln N  \nonumber \\
&&-\frac{\sqrt{2}}N\omega ^{(AB)}{}_{EC}\stackrel{\circ }{\nabla }%
_{(AB)}N^{CE},
\end{eqnarray}
\begin{eqnarray}
{\cal H}_{AB} &=&\sqrt{2}\sigma [-\stackrel{\circ }{\nabla }_{(CD)}\omega
^{(CD)}{}_{AB}+2\omega ^{(CD)}{}_{BE}\omega _{(CD)A}{}^E]  \nonumber \\
&&-\zeta _b{}^C\widetilde{p}^{bD}\omega _{(AB)CD},
\end{eqnarray}
and 
\begin{equation}
\widetilde{B}^{(AB)}=-\frac 12(N\sigma \omega ^{(CB)A}{}_C+N\sigma \omega
^{(CA)B}{}_C)-\sqrt{2}\sigma N^{CD}\omega ^{(AB)}{}_{DC}.
\end{equation}
Here $\widetilde{B}^{(AB)}$ is just the self-dual part of the boundary term $%
\widetilde{B}^\mu $ given by Nester [5].

By following the Dirac constraint analysis we find that the theory has the
same constraint structure. There are only two constraints, the scalar
constraint 
\[
{\cal H}_{\perp }=0, 
\]
and the vector constraint 
\[
{\cal H}_{AB}=0. 
\]
The phase space ($\Gamma _{TG},\Omega _{TG}$) of the teleparallel gravity is
coordinatized by the pair ($\zeta _{bD},\widetilde{p}^{bD}$) and has
symplectic structure 
\begin{equation}
\Omega _{TG}=\int_\Sigma d\widetilde{p}^{bD}\wedge d\zeta _{bD}.
\end{equation}
By constructing the constraint functions by smearing ${\cal H}_{\perp }$\
and ${\cal H}_i$\ with test fields $N$\ and $N^i$\ on $\Sigma $ following
the approach of Ashtekar [8] 
\begin{eqnarray*}
C(N) &=&\int_\Sigma N{\cal H}_{\perp }, \\
C(\overrightarrow{N}) &=&\int_\Sigma N^{AB}{\cal H}_{AB},
\end{eqnarray*}
we find that in the case $\partial _iN^i=0$, the constraint algebra is given
by 
\begin{equation}
\{C(N),C(M)\}=C({\cal L}_{\overrightarrow{t}}M)-C({\cal L}_{\overrightarrow{t%
}}M),
\end{equation}
\begin{equation}
\{C(\overrightarrow{N)},C(M)\}=C({\cal L}_{\overrightarrow{N}}M),
\end{equation}
and 
\begin{equation}
\{C(\overrightarrow{N)},C(\overrightarrow{M})\}=C({\cal L}_{\overrightarrow{N%
}}\overrightarrow{M})=C([\overrightarrow{N},\overrightarrow{M}]).
\end{equation}
These equations indicate that the constraint algebra is closed and the
constraints $C(N)$ and $C(\overrightarrow{N)}$ are first class, which is
very similar to the case in general relativity. The first class constraints $%
C(N)$ and $C(\overrightarrow{N)}$ generate the corresponding gauge
transformations, the spacetime translations. We have shown that the
constraint algebra of the teleparallel gravity has the same structure as
that of general relativity.

\section{Witten-Nester gauge conditions}

Introducing the Lorentz covariant derivative of $e^I{}_\nu $ by 
\[
\triangledown _\mu e^I{}_\nu =\partial _\mu e^{I{}}\!_\nu +\omega _\mu
\!^I\!_Je^J\!_\nu , 
\]
then we have 
\[
\omega _\mu \!^\rho \!_\nu =e_I{}^\rho e^J{}_\nu \omega _\mu
\!^I\!_J=e_I{}^\rho (\triangledown _\mu e^I{}_\nu -\partial _\mu
e^{I{}}\!_\nu ). 
\]
Using the dyad 
\begin{equation}
\zeta _{0A}=o_A,\zeta _{1A}=\iota _A,\zeta ^{A0}=-\iota ^A,\zeta ^{A1}=o^A,
\end{equation}
and supposing 
\begin{equation}
o{}^A=\frac 1\chi \lambda {}^A,\iota {}^A=\frac 1\chi \lambda {}^{\dagger A},
\end{equation}
one obtains 
\begin{eqnarray}
\omega _{CDAB} &=&\zeta _{aA}(\triangledown _{CD}\zeta ^a{}_B-\partial
_{CD}\zeta ^a{}_B)  \nonumber \\
&=&\frac 1{\chi ^2}(\lambda {}_A^{\dagger }\nabla _{CD}\lambda _B-\lambda
_A\triangledown _{CD}\lambda {}_B^{\dagger }  \nonumber \\
&&-\lambda {}_A^{\dagger }\partial _{CD}\lambda _B+\lambda _A\partial
_{CD}\lambda {}_B^{\dagger }),
\end{eqnarray}
and then 
\begin{eqnarray*}
&&\omega ^{(CB)A}{}_C+\omega ^{(CA)B}{}_C \\
&=&\frac 1{\chi ^2}(\lambda {}^{\dagger A}\triangledown ^{(CB)}\lambda
_C-\lambda ^A\triangledown ^{(CB)}\lambda {}_C^{\dagger }+\lambda
{}^{\dagger B}\triangledown ^{(CA)}\lambda _C-\lambda ^B\triangledown
^{(CA)}\lambda {}_C^{\dagger } \\
&&-\lambda {}^{\dagger A}\partial ^{(CB)}\lambda _C+\lambda ^A\partial
^{(CB)}\lambda {}_C^{\dagger }-\lambda {}^{\dagger B}\partial ^{(CA)}\lambda
_C+\lambda ^B\partial ^{(CA)}\lambda {}_C^{\dagger }).
\end{eqnarray*}
Suppose the spinors $\lambda ^A$ and its conjugate $\overline{\lambda }%
^{A^{\prime }}$ are the solutions of the Witten equation 
\begin{equation}
\triangledown _{(AB)}\lambda ^A=0,
\end{equation}
and 
\begin{equation}
\triangledown _{(A^{\prime }B^{\prime })}\overline{\lambda }^{A^{\prime }}=0.
\end{equation}
The later leads to 
\begin{equation}
\triangledown _{(AB)}\lambda ^{\dag A}=\frac 1{\sqrt{2}}K\lambda _B^{\dag }.
\end{equation}
Using (51) and (53) one can compute 
\begin{eqnarray}
&&\omega ^{(CB)A}{}_C+\omega ^{(CA)B}{}_C  \nonumber \\
&=&-\frac 1{\sqrt{2}\chi ^2}K(\lambda ^B\lambda {}^{\dagger A}+\lambda
^A\lambda {}^{\dagger B})  \nonumber \\
&&-\frac 1{\chi ^2}(\lambda {}^{\dagger A}\partial ^{(CB)}\lambda _C-\lambda
^A\partial ^{(CB)}\lambda {}_C^{\dagger }+\lambda {}^{\dagger B}\partial
^{(CA)}\lambda _C-\lambda ^B\partial ^{(CA)}\lambda {}_C^{\dagger }).
\end{eqnarray}
Introducing the triad on the spacelike hypersurface $\Sigma $: 
\begin{eqnarray}
e_1{}^{AB} &=&\frac 1{\sqrt{2}}(m^a-\overline{m}^a)=\frac 1{\sqrt{2}\chi ^2}%
(\lambda ^A\lambda ^B+\lambda ^{\dagger A}\lambda ^{\dagger B}),  \nonumber
\\
e_2{}^{AB} &=&\frac{-i}{\sqrt{2}}(m^a+\overline{m}^a)=\frac{-i}{\sqrt{2}\chi
^2}(\lambda ^A\lambda ^B-\lambda ^{\dagger A}\lambda ^{\dagger B}), 
\nonumber \\
e_3{}^{AB} &=&\frac 1{\sqrt{2}}(l^a-n^a)=\frac 1{\sqrt{2}\chi ^2}(\lambda
^A\lambda ^{\dagger B}+\lambda ^{\dagger A}\lambda ^B),
\end{eqnarray}
one can compute 
\[
\partial ^{(AB)}e_{1AB}=-2\chi ^{-1}\partial ^{(AB)}\chi e_{1AB}+\frac{\sqrt{%
2}}{\chi ^2}(\lambda _A\partial ^{(AB)}\lambda _B+\lambda _A^{\dagger
}\partial ^{(AB)}\lambda _B^{\dagger }), 
\]
\begin{eqnarray*}
(\omega ^{(CB)A}{}_C+\omega ^{(CA)B}{}_C)e_{1AB} &=&-\frac{\sqrt{2}}{\chi ^2}%
(\lambda _A\partial ^{(AB)}\lambda _B+\lambda _A^{\dagger }\partial
^{(AB)}\lambda _B^{\dagger }) \\
&=&-2\chi ^{-1}\partial ^{(AB)}\chi e_{1AB}-\partial ^{(AB)}e_{1AB} \\
&=&-\partial _1\ln \chi ^2-\partial ^{(AB)}e_{1AB}.
\end{eqnarray*}
By the same way one gets 
\[
(\omega ^{(CB)A}{}_C+\omega ^{(CA)B}{}_C)e_{2AB}=-\partial _2\ln \chi
^2-\partial ^{(AB)}e_{2AB}, 
\]
\[
(\omega ^{(CB)A}{}_C+\omega ^{(CA)B}{}_C)e_{3AB}=-\partial _3\ln \chi
^2-\partial ^{(AB)}e_{3AB}+K. 
\]
Then we have 
\begin{eqnarray}
q_I &=&\omega _{jI}^{+}{}^j=(\omega ^{(CB)A}{}_C+\omega ^{(CA)B}{}_C)e_{IAB}
\nonumber \\
&=&-\partial _I\ln \chi ^2-\partial ^{(AB)}e_{IAB}+\delta _{I3}K,  \nonumber
\\
(I &=&1,2,3).
\end{eqnarray}
This is just the Nester gauge condition with a correct term $\partial
^{(AB)}e_{IAB}$.

\section{Witten identity and positivity of gravitational energy}

Using (53) one can compute 
\begin{eqnarray*}
-N^j\omega ^{+i}{}_{\perp j} &=&-\sqrt{2}N^{CD}\omega ^{(AB)}{}_{DC} \\
&=&-\frac{\sqrt{2}}{\chi ^2}N^{CD}[\lambda _C\nabla ^{(AB)}\lambda
_D^{\dagger }-\lambda {}_C^{\dagger }\nabla \partial ^{(AB)}\lambda _D] \\
&&+\frac{\sqrt{2}}{\chi ^2}N^{CD}[\lambda _C\partial ^{(AB)}\lambda
_D^{\dagger }-\lambda {}_C^{\dagger }\partial ^{(AB)}\lambda _D].
\end{eqnarray*}
Supposing 
\begin{equation}
N^{CD}=-\frac 1{\sqrt{2}}\lambda ^{(\dagger C}\lambda ^{D)},
\end{equation}
we have 
\begin{eqnarray*}
-N^j\omega ^{+i}{}_{\perp j} &=&\sqrt{2}N^{CD}n_C{}^E\omega ^{(AB)}{}_{DE} \\
&=&\lambda ^{\dagger C}\nabla ^{(AB)}\lambda _C-\lambda ^{\dagger C}\partial
^{(AB)}\lambda _C.
\end{eqnarray*}
If we choose 
\begin{equation}
N=\lambda _A\lambda ^{\dagger A}=\chi ^2,
\end{equation}
the integral of the boundary term (46) reads 
\begin{eqnarray}
&&%
\displaystyle \oint %
\limits_S\widetilde{B}^{(AB)}dS_{AB}  \nonumber \\
&=&-\frac 12%
\displaystyle \oint %
\limits_S\sigma (N\omega ^{(CB)A}{}_C+N\omega ^{(CA)B}{}_C)dS_{AB}  \nonumber
\\
&&+2%
\displaystyle \oint %
\limits_S\sigma N^{CD}n_C{}^E\omega ^{(AB)}{}_{DE}dS_{AB}  \nonumber \\
&=&%
\displaystyle \oint %
\limits_S\sigma \chi ^2(\partial _I\ln \chi ^2+\partial ^{(AB)}e_{IAB})dS^I+%
\displaystyle \oint %
\limits_S\sigma KdS^3  \nonumber \\
&&%
\displaystyle \oint %
\limits_S\sigma \{\lambda ^{\dagger C}\nabla ^{(AB)}\lambda _C-\lambda
^{\dagger C}\partial ^{(AB)}\lambda _C\}dS_{AB}.
\end{eqnarray}
Using the Witten identity [19] 
\begin{eqnarray}
&&\oint_S\sigma \lambda ^{\dagger A}\nabla _i\lambda _AdS^i  \nonumber \\
&=&2\int_\Sigma {}\sigma (\nabla ^{(BC)}\lambda ^A)^{\dagger }(\nabla
_{(BC)}\lambda _A)dV+  \nonumber \\
&&4\pi G\int_\Sigma \sigma \lambda ^{\dagger A}(T_{00}\lambda _A+\sqrt{2}%
T_{0AB}\lambda ^B)dV,
\end{eqnarray}
one find 
\begin{eqnarray}
&&%
\displaystyle \oint %
\limits_S\widetilde{B}^{(AB)}dS_{AB}  \nonumber \\
&=&\frac 1{2\sqrt{2}}%
\displaystyle \oint %
\limits_S\sigma K(\lambda ^B\lambda {}^{\dagger A}+\lambda ^A\lambda
{}^{\dagger B})dS_{AB}  \nonumber \\
&&+\frac 12%
\displaystyle \oint %
\limits_S\sigma (\lambda {}^{\dagger A}\partial ^{(CB)}\lambda _C-\lambda
^A\partial ^{(CB)}\lambda {}_C^{\dagger }+\lambda {}^{\dagger B}\partial
^{(CA)}\lambda _C  \nonumber \\
&&-\lambda ^B\partial ^{(CA)}\lambda {}_C^{\dagger }-\lambda ^{\dagger
C}\partial ^{(AB)}\lambda _C)dS_{AB}  \nonumber \\
&&+2\int_\Sigma {}\sigma (\nabla ^{(BC)}\lambda ^A)^{\dagger }(\nabla
_{(BC)}\lambda _A)dV  \nonumber \\
&&+4\pi G\int_\Sigma \sigma \lambda ^{\dagger A}(T_{00}\lambda _A+\sqrt{2}%
T_{0AB}\lambda ^B)dV.
\end{eqnarray}
which leads the positivity of the gravitational energy in the asymptotically
flat boundary condition in space infinity. It is to be noted that we do not
use Nester's gauge $N^{AB}=0$. Instead we suppose the equation (60), which
means that Nester's proof is extended to the case including momentum. The
Nester gauge condition plays role only in the lapse part of the boundary
term, while the Witten equation plays roles not only in the lapse part but
also in the shift part of the boundary term in the proof of the positive
energy theorem.

\section{conclusions}

A self-dual teleparallel gravity is developed. Its Lagrangian is equivalent
the Ashtekar Lagrangian. The basic dynamic variables are the dyad spinors $%
\zeta _{aA}$. The Ashtekar connection appears in the canonical momentum
conjugate $\widetilde{p}^{aA}$ to $\zeta _{aA}$. In the Hamiltonian
formulation of this theory the Nester gauge condition can be derived from
the Witten equation directly and a new expression for the boundary term
which is the self-dual part of the Nester boundary term is obtained. Using
this expression the proof of the positive energy theorem by Nester et al can
be extended to a case including momentum.

{\bf ACKNOWLEDGMENT}$\;$

This work is completed when G. Y. C. visits Institute of Applied
Mathematics, Academy of Mathematics and System Science, Chinese Academy of
Science and supported by the National Science Foundation Grants of China No
10175032. and the Science Foundation Grants of the Education Department of
Liaoning Province No. 20041011.


\begin{references}
\bibitem{}  V. C. de Andrade, L. C. T. Guillen, and J. G. Pereira, Phys.
Rev. Lett.{\it \ }{\bf 84}, 4533 (2000); V. C. de Andrade, and J. G.
Pereira, Int. J. Mod. Phys. {\bf D 8}, 141(1999); Gen. Rel. Grav. {\bf 30},
263 (1998); Phys. Rev. {\bf D 56}, 4689 (1997); M. Calcada and J. G.
Pereira, Phys. Rev. {\bf D 66}, 044001 (2002).

\bibitem{}  J. W. Maluf, and J. F. da Rocha-Neto, Gen. Rel. Grav. {\bf 31},
173 (1999); J. W.Maluf, E. F. Matins, and A. Kneip, J. Math. Phys. {\bf 37},
6302 (1996); J. W. Maluf, Gen. Rel. Grav. {\bf 28}, 1361 (1996); J. Math.
Phys. {\bf 37}, 6293 (1996); J. Math. Phys. {\bf 36}, 4242 (1995); J. Math.
Phys. {\bf 35}, 335 (1994);

\bibitem{}  M. Blagojevic and I. A. Nikolic, Phys. Rev{\it .} {\bf D 62},
024021 (2000); M. Blagojevic and M. Vasilic, Phys. Rev. {\bf D 64}, 044010
(2001); G. G. L. Nashed, Phys. Rev. {\bf D 66}, 064015 (2002).

\bibitem{}  U. Muench, F. Gronwald, and F. W. Hehl, Gen. Rel. Grav{\it . }%
{\bf 30}, 933(1998).

\bibitem{}  C. C. Chang, J. M. Nester, and C. M. Chen, Phys. Rev. Lett. {\bf %
83}, 1897 (1999); R. S. Tung, and J. M. Nester, Phys. Rev.{\it \ }{\bf D 60}%
, 021501 (1999); J. M.\ Nester, and H. J. Yo, Chin. J. Phys.{\it \ }{\bf 37}%
, 113 (1999); J. M. Nester,{\it \ }Int. J. Mod. Phys.{\it \ }{\bf A4}, 1755
(1989).

\bibitem{}  Y. M. Cho, Phys. Rev{\it . }{\bf D 14}, 2521 (1976).

\bibitem{}  E. W. Mielke, {\it Phys. Lett.} {\bf 251A}, 349 (1999); Phys.
Lett. {\bf 149A}, 345 (1992); Ann. Phys (N. Y. ) {\bf 219}, 78 (1992); Phys.
Rev.{\it \ }{\bf D42}, 3388 (1990).

\bibitem{}  A. Ashtekar, Phys. Rev. Lett. {\bf 57}, 2244 (1986); Phys. Rev%
{\it .} {\bf D36}, 1587 (1987); A. Ashtekar, J. D. Romano, and R. S. Tate,
Phys. Rev. {\bf D 40}, 2572 (1989).

\bibitem{}  D. M. Kerrick, Phys. Rev. Lett. {\bf 75}, 2074 (1995); F.
Wilczek, Phys. Rev. Lett. {\bf 80}, 4851 (1998).

\bibitem{}  T. Appelquist, A. Chodos, and P. G. O. Freund, {\it Modern
Kaluza-Klein Theories }(Addison-Wesley, Reading, MA, 1987).

\bibitem{}  V. C. de Andrade, L. C. T. Guillen, and J. G. Pereira, {\it %
Phys. Rev.} {\bf D 61}, 084031 (2000); A. L. Barbosa, L. C. T. Guillen, and
J. G. Pereira, Phys. Rev. {\bf D 66}, 064028 (2002).

\bibitem{}  C. Moller,{\it \ }Mat. Fys. Skr. Dan. Vidensk. Selsk.{\it \ }%
{\bf 1}, No. 10 (1961); Ann. Phys. {\bf 12},118 (1961).

\bibitem{}  R, Penrose Proc. R. Soc. London {\bf A 381}, 53 (1982).

\bibitem{}  J. D. Brown, and J. W. York, {\it Phys. Rev.} {\bf D 47, }1407
(1993) and references therein.

\bibitem{}  R. S. Tung and T. Jacobson, Class. Quantum Grav. {\bf 12}, L51
(1995); R. S. Tung, Phys. Lett. {\bf A264}, 341 (2000); C. M. Chen, J. M.
Nester, and R. S. Tung, arXiv:gr-qc/0209100v2.

\bibitem{}  R. S. Ward and R. O. Wells, {\it Twistor Geometry and Field
Theory }(Cambridge Univesity Press, Cambridge, England, 1990); L. J. Mason
and N. M. J. Woodhous, {\it Integrabality,} {\it Self-duality, and Twistor \
Theory }(Clarendon, Oxford, 1996)

\bibitem{}  J A. Nieto, O. Obregon, and J. Socorro, Phys. Rev. {\bf D 50},
R3583 (1994); J. A. Nieto, J. Socorro, and O. Obregon, Phys. Rev. Lett. {\bf %
76}, 3482 (1996).

\bibitem{}  G. Chee, Phys. Rev. {\bf D 54}, 6552 (1996); G. Y. Chee. Phys.
Rev. {\bf D 62}, 064013, 064014 (2000); Gen. Relat. Grav. {\bf 30}, 1735
(1998); X. Han and G. Y. Chee, Phys. Rev. {\bf D 62}, 084006 (2000); G. Y.
Chee and Y. H. Jia, Gen. Relat. Grav. {\bf 33}, 1953 (2001).

\bibitem{}  E. Witten, Commun. Math. Phys. {\bf 80}, 381 (1981)

\bibitem{}  J. Frauendiener, Class. Quantum. Grav. {\bf 8}, 1881 (1991); V.
Pelykh, J. Math. Phys. {\bf 41}, 5550 (2000).

\bibitem{}  F. W. Hehl, J. D. McCrea, E. W. Mielke, and Y.Ne'emann, Phys.
Rep. {\bf 258}, 27 (1995).

\bibitem{}  K. Hayashi, and T. Shirafuji, Phys. Rev.{\it \ }{\bf D19}, 3524
(1979).

\bibitem{}  S. Kobayashi, and K. Nomizu, {\it Foundation of Differential
Geometry} Vol. I (New York: Interscience Publishers, 1963); T. Frankel, {\it %
The geometry of Physics, An Introduction }(Cambridge: Cambridge University
Press, 1997).

\bibitem{}  P. Sommers, J. Math. Phys. {\bf 21}, 2567 (1980).
\end{references}
\end{document}